\documentclass[notitlepage,nofootinbib,preprintnumbers,aps,superscriptaddress,twocolumn]{revtex4-1}
\usepackage{amsmath,amssymb,natbib,bm,color,xcolor}
\usepackage{appendix}
\usepackage{graphicx}
\usepackage{slashed}
\usepackage{comment}

\begin{document}
\title{Probing High Reheating Temperatures by Direct Detection Experiments}
\author{Barmak Shams Es Haghi}
\email{barmak.shams@utah.edu}
\affiliation{Texas Center for Cosmology and Astroparticle Physics, Weinberg Institute for Theoretical Physics, Department of Physics, The University of Texas at Austin, Austin, TX 78712, USA}
\affiliation{Department of Physics and Astronomy, University of Utah, Salt Lake City, UT, 84112, USA}

\begin{abstract}
We argue that the benchmark freeze-in dark matter (DM) scenario for direct detection experiments, in which a DM candidate interacts with the Standard Model (SM) through an ultralight dark photon, becomes sensitive to the visible sector reheating temperature if it is sufficiently high $(\gtrsim 10^{15}\,{\rm GeV})$. At such temperatures, the irreducible ultraviolet (UV) freeze-in production of DM through graviton exchange becomes important and must be combined with the infrared (IR) freeze-in yield mediated by the dark photon. As long as gravitationally produced DM does not equilibrate through annihilation into dark photons and the subsequent formation of a dark thermal bath, it retains information about the reheating phase. Including this gravitational contribution relaxes the required DM–SM portal coupling and allows smaller values than those that would match the observed relic abundance through IR freeze-in alone. Since current direct detection experiments have excluded the benchmark freeze-in model over a wide range of DM masses, they are now effectively probing high reheating temperatures and the gravitational freeze-in of DM.
\end{abstract}
 
\maketitle
\textbf{\textit{Introduction.}} 
Despite extensive searches for dark matter (DM) particles, all results to date have been null, and the true nature of DM remains unknown. Among these efforts, direct detection experiments have achieved unprecedented sensitivity to DM–nucleon interactions~\cite{PandaX-II:2017hlx,XENON:2018voc,LZ:2022lsv,LZ:2024zvo} and have also made remarkable progress in probing DM–electron interactions in the sub-GeV mass range~\cite{SENSEI:2020dpa,PandaX:2022xqx,DarkSide:2022knj,DAMIC-M:2023gxo,DAMIC-M:2023hgj,DAMIC-M:2025luv,Zhang:2025ajc}. A benchmark and minimal model that has been a primary target of direct detection experiments in the MeV-TeV mass range consists of a DM candidate that interacts with the Standard Model (SM) through an ultralight dark photon kinetically mixed with the SM photon~\cite{Hall:2009bx,Chu:2011be,Essig:2011nj}. Because the mixing parameter is very small, DM is produced through the dark photon via the freeze-in mechanism: DM particles are generated gradually during the thermal history of the Universe without ever reaching thermal equilibrium with the bath~\cite{Hall:2009bx}. Since production is suppressed by a tiny dimensionless coupling, the dominant contribution to the DM abundance arises when the temperature of the SM bath is close to the DM mass. This is known as infrared (IR) freeze-in, because the yield depends on the DM mass rather than the temperature of the bath at early times, assuming the reheating temperature is much larger than the DM mass~\cite{Hall:2009bx}. For a fixed DM mass, the present-day abundance uniquely determines the mixing parameter that couples DM to the SM. This picture changes if one assumes that the reheating temperature of the visible sector, $T_{\rm rh}$, is below the DM mass but still above the MeV scale~\cite{Boddy:2024vgt}. In this case, most DM production occurs at the reheating temperature, and because the production rate is Boltzmann suppressed, a larger coupling is required to reproduce the observed relic abundance. Recent direct detection results have not only probed the low–reheating temperature version of this freeze-in benchmark model, but have also excluded it over a wide range of DM masses when $T_{\rm rh}\gg100\,{\rm TeV}$~\cite{DAMIC-M:2025luv,Bernal:2024ndy}. The conclusion is that, across a broad mass range, this DM candidate can only constitute a subdominant fraction of the total DM density.

In this {\it Letter}, we highlight that the thermal bath inevitably produces an additional, irreducible population of DM particles through gravitational interactions, mediated by the exchange of a massless spin-2 graviton~\cite{Garny:2015sjg,Tang:2017hvq,Garny:2017kha,Bernal:2018qlk}. Because gravitational interactions are suppressed by the Planck scale, this production channel is most efficient at the highest temperatures reached by the bath, and is therefore known as ultraviolet (UV) freeze-in~\cite{Elahi:2014fsa}.
Although for DM in the MeV–TeV mass range and for moderately high reheating temperatures this contribution is negligible, it can become significant at very high reheating temperatures and should be included. As long as the gravitationally produced DM does not equilibrate through annihilation into dark photons and the formation of a dark thermal bath, it retains information about the reheating stage. Adding the gravitationally produced component to the IR freeze-in yield, mediated by the dark photon, reduces the portal coupling between DM and the SM, thereby opening up the parameter space. By treating the reheating temperature as a free parameter and including the gravitational contribution, the benchmark model can account for the full DM relic abundance, provided the reheating temperature is sufficiently high. Consequently, current direct detection experiments are already probing the DM candidate together with the reheating scale and the gravitational production of DM.

\textbf{\textit{Minimal Dark Matter Freeze-in Model.}} The benchmark model that we consider includes a Dirac fermion DM particle charged under a dark $U(1)'$ gauge field which mixes kinetically with the SM hypercharge:
\begin{equation}
    \mathcal{L}\supset \bar{\chi}(i\slashed{D}-m_\chi)\chi-\frac{\epsilon}{2}F^{\mu\nu}_Y F'_{\mu\nu},
\end{equation}
with $\slashed{D}=\partial_\mu+ie'A'_\mu$, where $A'_\mu$ is the dark photon, $e'$ is the DM charge and $m_\chi$ its mass, $F^{\mu\nu}_Y$ and $F'_{\mu\nu}$ are  the field strengths of the SM hypercharge and the dark photon, respectively, and $\epsilon$ is the mixing parameter. The dark photon has a negligible mass, $m_{\gamma'}$, which enhances the direct detection cross section at low momentum transfer and compensates for the small coupling to the SM in direct detection searches. We assume that $m_{\gamma'}\lesssim 10^{-15}\,{\rm eV}$ to remain consistent with constraints from black hole superradiance~\cite{Baryakhtar:2017ngi,Siemonsen:2022ivj} and COBE/FIRAS~\cite{Fixsen:1996nj,Caputo:2020bdy}.

\textbf{\textit{Dark Matter Abundance: IR and UV Freeze-in.}}
The DM abundance is established by freeze-in from the SM thermal bath through two channels: dark photon exchange and graviton exchange. As long as the thermal bath remains in equilibrium (with particles scattering continuously), DM is produced through these two channels. Because each channel is suppressed for different reasons, their production histories peak at different times: graviton-mediated production is largest at the reheating temperature, while dark photon-mediated production peaks when the bath temperature drops to around the DM mass.

It is customary to introduce the portal coupling $\kappa\equiv \epsilon \cos \theta_W \sqrt{\alpha'/\alpha}$ to capture all the parameters that connect the thermal bath to the DM, where $\alpha (\alpha')$ is the electromagnetic (dark) fine structure constant, and $\theta_W$ is  the Weinberg angle~\cite{Chu:2011be}. All the particles within the SM that can produce $\bar\chi\chi$ at tree level contribute in IR freeze-in~\cite{Essig:2011nj,Chu:2011be} with a production rate proportional to $\kappa^2$ (for a revised and corrected calculation of IR freeze-in, see Ref.~\cite{Bhattiprolu:2023akk}). 

Gravitational freeze-in can be understood by expanding the metric around flat space-time, $g_{\mu\nu}\simeq \eta_{\mu\nu}+h_{\mu\nu}/M_{\rm Pl}$ where $M_{\rm Pl}\simeq2.4\times 10^{18}\,{\rm GeV}$ is the reduced Planck mass. Then equivalence principle introduces a coupling between the SM and $\chi$ through the energy-momentum tensors:
\begin{equation}
    \mathcal{L}\supset \frac{1}{2M_{\rm Pl}}h_{\mu\nu}\left(T^{\mu\nu}_{\rm SM}+T_\chi^{\mu\nu}\right),
\end{equation}
which in turn leads to the annihilation of SM particles into DM  through $s$-channel graviton exchange~\cite{Garny:2015sjg,Tang:2017hvq,Garny:2017kha,Bernal:2018qlk}. All the SM fields contribute to DM production through the gravitational channel, with their energy-momentum tensors, which depend on the spin of each field.
The evolution of the UV freeze-in yield of the DM is given by:
\begin{equation}
    \frac{dY^{\rm UV}_\chi}{dT}=\frac{-n^2_{\chi,{\rm eq}}\langle\sigma v\rangle_{\rm UV}}{H(T)Ts(T)},
\end{equation} 
where
$H(T)=(\pi/3\sqrt{10})g_\star^{1/2}(T)T^2/M_{\rm Pl}$ is the Hubble expansion rate during radiation domination, $s(t)=(2\pi^2/45)g_{\star,S}(T)T^3$ is the entropy density of the thermal bath, and $n^2_{\chi,{\rm eq}}\langle\sigma v\rangle_{\rm UV}=c_{1/2}T^8/M^4_{\rm Pl}$ with prefactor $c_{1/2}\simeq 1.1 \times 10^{-3}$~\cite{Bernal:2018qlk} which represents the contribution of all SM fields with different spins to the production of a Dirac fermion through $s$-channel graviton exchange. The relic abundance of DM from UV freeze-in, starting from a negligible initial value, is given by:
\begin{equation}
    \Omega^{\rm UV}_\chi h^2\simeq 4.5\times 10^{-3}\left(\frac{m_\chi}{10^5\,{\rm GeV}}\right)\left(\frac{T_{\rm rh}}{10^{15}\,{\rm GeV}}\right)^3,
\end{equation}
for $g_\star=g_{\star,S}=106.75$. Therefore, gravitationally produced DM with a mass of $100\,{\rm TeV}$, generated at reheating temperatures of order 
$10^{15}\,{\rm GeV}$, can constitute a few percent of the observed DM abundance, provided it does not equilibrate and become part of a dark thermal bath.

Gravitationally produced DM may annihilate into dark photons and form a dark thermal bath. Thermalization in the dark sector would erase the connection to the reheating stage. To prevent this, we require that the annihilation rate remain below the Hubble expansion rate. The ratio of annihilation rate to the Hubble rate, as long as DM particles are relativistic, is given by:
\begin{equation}
    \frac{\Gamma}{H}\sim \frac{n_\chi\langle\sigma v\rangle_{\bar\chi\chi\rightarrow\gamma'\gamma'}}{H}\sim \frac{n_{\chi,{\rm rh}}(4\pi\alpha'^2/T_{\rm rh}^2)}{(\pi/3\sqrt{10}) g_\star^{1/2}T_{\rm rh}^2/M_{\rm Pl}}\frac{a}{a_{\rm rh}},
\end{equation}
where $a$ is the scale factor. The $a/a_{\rm rh}$ factor account for redshifting: while the particles remain relativistic, the ratio grows. It reaches its maximum when particles become non-relativistic (which is when IR freeze-in becomes significant), i.e., at $T\sim m_\chi$. At that point $a/a_{\rm rh}$ can be as large as $T_{\rm rh}/m_\chi$. The maximum value of $\Gamma/H$ corresponds to the case where the entire DM relic abundance today, $\Omega_{\rm CDM}$, is produced at reheating, i.e., $n_{\chi,{\rm rh}}=(\rho_c/m_\chi)(s(T_{\rm rh})/s_0)\Omega_{\rm CDM}$ and when $a/a_{\rm rh}\sim T_{\rm rh}/m_\chi$:
\begin{eqnarray}
   \nonumber \frac{\Gamma}{H}&\lesssim&  \frac{8\sqrt{2}\pi^2}{3\sqrt{5}}\alpha'^2g^{1/2}_\star\Omega_{\rm CDM}\frac{\rho_c m_{\rm Pl}}{s_0 m_\chi^2}\\
    &\sim& 10^{-7}\left(\frac{\alpha'}{10^{-9}}\right)^2\left(\frac{1\,{\rm GeV}}{m_\chi}\right)^2,
\end{eqnarray} 
where $\rho_c$ is the critical energy density of the Universe, and $s_0$ is the entropy density of the Universe today. As we can see, for $\alpha'\lesssim 10^{-9}$ the annihilation rate is smaller than the Hubble rate over all the MeV-TeV mass range. For larger values of $\alpha'$, gravitationally produced DM can annihilate and create a dark bath with a temperature different from the one for the SM. In that case, the minimal freeze-in benchmark model can be used to probe the thermal history of a dark sector with a light mediator~\cite{Fernandez:2021iti}. We assume that $\alpha'$ is sufficiently small that gravitationally produced DM does not annihilate into dark photons and is therefore diluted solely by the expansion of the Universe and contributes directly to the final DM abundance. Consequently, the UV and IR contributions can be evaluated separately and then summed to obtain the total DM abundance: 
\begin{equation}
  \Omega_\chi(\kappa,m_\chi,T_{\rm rh})\equiv \Omega^{\rm IR}_\chi(\kappa,m_\chi,T_{\rm rh})+  \Omega^{\rm UV}_\chi(m_\chi, T_{\rm rh}).
\end{equation}
In the MeV-TeV mass range, the gravitational contribution to the relic abundance is negligible for reheating temperatures $T_{\rm rh}\lesssim 10^{15}\,{\rm GeV}$. Consequently, for $m_\chi<T_{\rm rh}\lesssim 10^{15}\,{\rm GeV}$, DM abundance is set entirely by IR freeze-in: $\Omega_\chi=\Omega^{\rm IR}_\chi(\kappa,m_\chi)$, which reproduces the traditional IR freeze-in result~\cite{Essig:2011nj,Chu:2011be}. For $T_{\rm rh}>10^{15}\,{\rm GeV}$, gravitational production becomes important. In this regime, the gravitationally produced component should be included, and therefore the IR freeze-in contribution is smaller than the total observed abundance of DM. This allows us to determine the modified coupling $\kappa(m_\chi, T_{\rm rh}>10^{15}\,{\rm GeV})$ required to match the observed relic abundance of DM today in terms of the IR-freeze-in coupling,  $\kappa_{\rm IR}\equiv\kappa(m_\chi,m_\chi<T_{\rm rh}<10^{15})$ which alone accounts for the full DM abundance in the absence of gravitational production:
\begin{equation}
    \frac{\kappa(m_\chi,T_{\rm rh}>10^{15}\,{\rm GeV})}{\kappa_{\rm IR}(m_\chi)}=\sqrt{1-\frac{\Omega^{\rm UV}_\chi(m_\chi, T_{\rm rh})}{\Omega_{\rm CDM}}}.
     \label{eq:kappa}
\end{equation}
It is evident that gravitational production at reheating temperature relaxes the required coupling between $\chi$ and the SM, while still allowing $\chi$ to account for the entire DM abundance.

\begin{figure}[t]
  \centering
    \includegraphics[width=\linewidth]{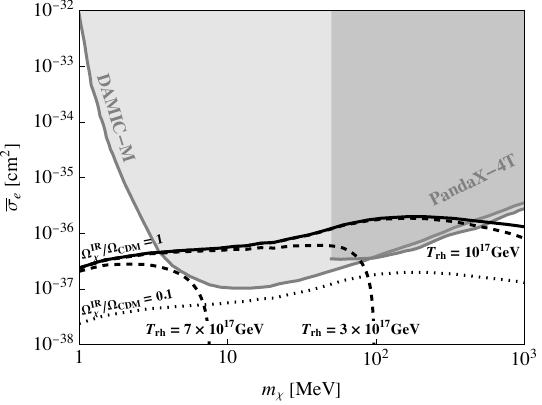}
  \caption{The DM-electron cross section for sub-GeV freeze-in DM, when gravitational production at reheating is included.
  For $m_\chi\ll T_{\rm rh}\ll10^{17}\,{\rm GeV}$, UV freeze-in production is negligible, and we obtain the standard IR freeze-in benchmark (black solid line), which is taken from Ref.~\cite{Essig:2011nj}. For $T_{\rm rh}\gtrsim 10^{17}\,{\rm GeV}$, the gravitational contribution increases and leads to a smaller DM-electron cross section. Consequently, the freeze-in scenario can extend to values below the black solid line (black dashed lines). The black dotted line corresponds to couplings that would produce $10\%$ of the final DM abundance via IR freeze-in (the remaining $90\%$ of the DM abundance is produced gravitationally). The gray shaded regions display exclusion limits from DAMIC-M~\cite{DAMIC-M:2025luv} and PandaX-4T~\cite{Zhang:2025ajc} experiments.}
\label{fig:1}
\end{figure}


\begin{figure}[htbp]
  \centering
    \includegraphics[width=\linewidth]{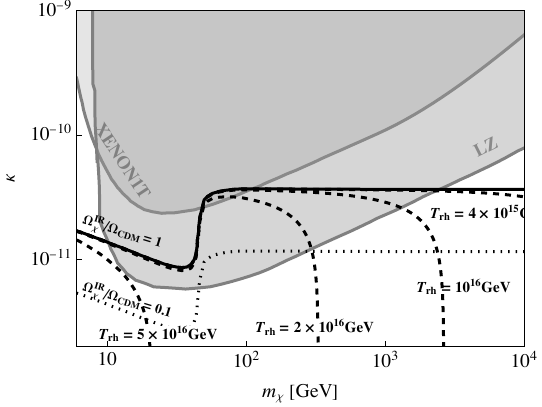}
  \caption{The DM-SM coupling for GeV-TeV freeze-in DM by including gravitational production at reheating.  For $m_\chi\ll T_{\rm rh}\ll10^{15}\,{\rm GeV}$, UV freeze-in production is insignificant, and we recover the standard IR freeze-in benchmark (black solid line), which is taken from Ref.~\cite{Bhattiprolu:2023akk}. For higher $T_{\rm rh}$, the gravitational contribution leads to a smaller coupling (black dashed lines). The black dotted line is described in the same way as in  Fig~\ref{fig:1}. We plot the current exclusion limits from XENON1T~\cite{XENON:2018voc} and LUX–ZEPLIN (LZ)~\cite{LZ:2024zvo} experiments, using the recast LZ constraints from Ref.~\cite{Bernal:2024ndy}, in gray shades.}
\label{fig:2}
\end{figure}

 \textbf{\textit{Impact on Direct Detection Searches.}}
The reduction of the DM–SM coupling due to the gravitationally produced component at reheating lowers the scattering rate of DM off electrons and nucleons in direct detection experiments, and in light of recent results, leads to interesting implications.

DM-electron scattering experiments express their results in term of DM-electron cross section defined by:
\begin{equation}
    \overline{\sigma}_e=\frac{16\pi \mu_{\chi e}^2\alpha^2\kappa^2}{(\alpha m_e)^4},
\end{equation}
where $\mu_{\chi e}$ and $m_e$ are the DM-electron reduced mass and the electron mass, respectively. 

In Fig.~\ref{fig:1}, we show the modified DM-electron cross section for sub-GeV freeze-in DM, when gravitational production at reheating is included. For $m_\chi\ll T_{\rm rh}\ll10^{17}\,{\rm GeV}$, UV freeze-in production is negligible for this mass range, and therefore we obtain the standard IR freeze-in benchmark (black solid line). By increasing the reheating temperature above $\sim 10^{17}\,{\rm GeV}$, the gravitational contribution increases and, according to Eq.~\eqref{eq:kappa}, leads to a smaller DM–electron cross section. Consequently, the freeze-in scenario is no longer restricted to the black solid line and can extend to values below it (black dashed lines). To clarify further, we also plot the cross sections corresponding to couplings that would produce $10\%$ of the final DM abundance via IR freeze-in (black dotted line). At the intersection of this line with the black dashed lines, the remaining $90\%$ of the DM abundance is produced gravitationally at the corresponding reheating temperature. The gray shaded regions illustrates exclusion limits from DAMIC-M~\cite{DAMIC-M:2025luv} and PandaX-4T~\cite{Zhang:2025ajc} experiments. Since the sensitivity of these experiments now extends below the standard IR freeze-in benchmark, they are already probing high reheating temperatures in the range $10^{17}\,{\rm GeV}\lesssim T_{\rm rh}\lesssim7\times 10^{17}\,{\rm GeV}$, and gravitational production of DM.

In Fig.~\ref{fig:2}, we display the modified DM-SM coupling for GeV-TeV freeze-in DM by including gravitational production at reheating.  We
show the nuclear-recoil exclusion limits from XENON1T~\cite{XENON:2018voc,Hambye:2018dpi,Bhattiprolu:2023akk} and LUX-ZEPLIN (LZ) experiments~\cite{LZ:2024zvo,Bernal:2024ndy} in gray shades. For GeV-TeV freeze-in DM, currently high reheating temperatures in the range $4 \times 10^{15}\,{\rm GeV}\lesssim T_{\rm rh}\lesssim 5\times 10^{16}\,{\rm GeV}$ are being probed.

It is worth mentioning that the new technique for sub-GeV DM direct detection that utilizes neutrino observatories~\cite{Leane:2025efj} can also turn these searches into probes of high reheating temperatures in the early Universe.
                                                 
\textbf{\textit{High Reheating Temperatures, Inflation and Alternatives.}}
So far, we have considered the reheating temperature as a free parameter 
which can be probed by current direct detection experiments in the range $\sim10^{15}\,{\rm GeV}-10^{17}\,{\rm GeV}$ within the context of the minimal freeze-in model. 
Although consistency with the successful predictions of big bang nucleosynthesis (BBN) imposes a lower limit of roughly $\sim 5\,{\rm MeV}$ on the reheating temperature~\cite{Hasegawa:2019jsa,Kawasaki:2000en}, there is no upper bound below the Planck scale. Avoiding the production of dangerous relics imposes certain model-dependent upper bounds on the reheating temperature. For instance, in the context of grand unified theories (GUTs), the reheating temperature should remain below the GUT scale ($\sim 10^{16}\,{\rm GeV}$) to avoid producing heavy magnetic monopoles~\cite{1978PhLB...79..239Z,Preskill:1979zi}. In supergravity theories, avoiding gravitino overproduction similarly leads to an upper bound on the reheating temperature~\cite{1993PhLB..303..289M,Asaka:2000zh,Pradler:2006hh,Steffen:2008bt,Covi:2010au}, e.g., $\sim10^9\,{\rm GeV}$~\cite{Heisig:2013sva}. However, in the absence of any evidence supporting these specific models, the reheating temperature can be treated as a free parameter and should be probed experimentally. 
Inflation~\cite{Starobinsky:1980te,Guth:1980zm,Linde:1981mu,Albrecht:1982wi}, widely regarded as the leading solution to the horizon, flatness, and other fundamental problems of Standard Big Bang cosmology, resolves these issues through a period of accelerated expansion preceding radiation domination. In slow-roll inflationary models, the energy scale of the inflation, $V^{1/4}$, is related to the amplitude of primordial scalar perturbations, $A_S$, and the tensor-to-scalar ratio, $r$, by $V^{1/4}=(3\pi^2 r A_S/2)^{1/4}M_{\rm Pl}$.
The non-observation of inflationary tensor fluctuations allows cosmic microwave background (CMB) measurements by the BICEP/Keck and {\it Planck} collaborations to place an upper bound on $r$ and consequently on the energy scale of inflation: $V^{1/4}\lesssim 1.6\times 10^{16}\,{\rm GeV}$~\cite{Planck:2018jri}. Assuming instantaneous reheating and $g_\star=106.75$, this corresponds to an upper bound on the reheating temperature: $T_{\rm rh}\lesssim 6.6\times 10^{15}\,{\rm GeV}$. Therefore, achieving a reheating temperature above $~\sim 10^{16}\,{\rm GeV}$ via instantaneous reheating would be in tension with slow-roll inflationary cosmology. In contrast, alternative scenarios such as ekpyrotic~\cite{Khoury:2001wf,Khoury:2001bz,Khoury:2001zk}, cyclic~\cite{Steinhardt:2001st}, and string gas cosmology~\cite{Brandenberger:1988aj,Tseytlin:1991xk,Kripfganz:1987rh} predict an extremely suppressed primordial tensor spectrum compared to inflation~\cite{Boyle:2003km,Brandenberger:2006xi}. Consequently, in these models, the non-observation of tensor modes does not constrain the reheating temperature.

All results in this study assume instantaneous reheating, where the maximum radiation temperature equals the reheating temperature. If one instead considers gravitational production of DM during reheating or from a subdominant thermal bath present during inflation, the DM yield could be enhanced by a few orders of magnitude compared to UV freeze-in in a radiation-dominated era with the same reheating temperature~\cite{Garcia:2017tuj,Chen:2017kvz,Bernal:2019mhf,Garcia:2018wtq,Freese:2024ogj,Wang:2025duy}. This enhancement would, optimistically, allow gravitationally produced DM to be generated at reheating temperatures up to roughly an order of magnitude lower than those considered here, bringing them closer to being compatible with slow-roll inflationary cosmology. We leave a careful study of these possibilities for future work. 

\textbf{\textit{Gravitational Wave Signal and Multimessenger Cosmology.}}
There exists an irreducible stochastic gravitational wave (GW) background produced by the SM thermal bath via graviton emission during particle scatterings. This background peaks at high frequencies $(\sim 100\,{\rm GHz})$, with an amplitude that depends on the reheating temperature~\cite{Ghiglieri:2015nfa,Ghiglieri:2020mhm,Ringwald:2020ist}. Despite recent interest in and attempts to detect high-frequency GWs~\cite{Aggarwal:2020olq,Aggarwal:2025noe}, observing those from the thermal bath of the early Universe remains extremely challenging; only very high reheating temperatures, close to the Planck scale, could make a future detection possible~\cite{Ringwald:2020ist}. Nevertheless, resonant cavity designs may achieve sufficient sensitivity in the terahertz band, providing a promising avenue to probe these GWs and explore the highest temperature of the Universe~\cite{Herman:2022fau}. Furthermore, the predicted GW signal could be enhanced if contributions from reheating~\cite{Bernal:2024jim} or a thermal bath during inflation~\cite{Montefalcone:2025gxx} are included. In combination, high-frequency GW searches and the DM direct detection strategy presented here offer a compelling and complementary multimessenger approach to exploring new physics and the early stages of the Universe.

\textbf{\textit{Discussion.}}
In this ${\it Letter}$, we have investigated the impact of high reheating temperatures on the benchmark freeze-in DM scenario. We showed that for MeV-TeV DM mas range, at very high reheating temperature $( T_{\rm rh}> 10^{15}\,{\rm GeV})$, the irreducible ultraviolet (UV) freeze-in production of DM
through graviton exchange becomes important. As long as gravitationally produced DM does not equilibrate through annihilation into dark photons and the subsequent formation of a dark thermal bath, it retains information about the reheating stage and contribute directly to the final abundance of DM. Because of this gravitational contribution, the IR freeze-in component must remain below the total observed DM abundance, implying a smaller DM–SM coupling than would be required to match the relic abundance through IR freeze-in alone. Since current direct detection experiments have already excluded the benchmark freeze-in model over a wide range of DM masses, they are now effectively probing the reheating stage and the gravitational freeze-in production of DM. Within the minimal freeze-in framework, a potential future signal below the traditional freeze-in benchmark would constitute a direct probe of the reheating phase, the highest temperature attained by the early Universe, and gravitational production of DM.

\textbf{\textit{Acknowledgments.}} We are grateful to Sonia Paban for insightful correspondence on reheating scales in inflationary cosmology and its alternatives. We also thank Evangelos I. Sfakianakis for helpful discussions about reheating in inflationary cosmology. We are particularly grateful to Pearl Sandick for her continuous support and encouragement. The work of B.S.E. is supported in part by the National Science Foundation
grant PHY-2412834.
\bibliography{main}{}

\begin{thebibliography}{67}%
\makeatletter
\providecommand \@ifxundefined [1]{%
 \@ifx{#1\undefined}
}%
\providecommand \@ifnum [1]{%
 \ifnum #1\expandafter \@firstoftwo
 \else \expandafter \@secondoftwo
 \fi
}%
\providecommand \@ifx [1]{%
 \ifx #1\expandafter \@firstoftwo
 \else \expandafter \@secondoftwo
 \fi
}%
\providecommand \natexlab [1]{#1}%
\providecommand \enquote  [1]{``#1''}%
\providecommand \bibnamefont  [1]{#1}%
\providecommand \bibfnamefont [1]{#1}%
\providecommand \citenamefont [1]{#1}%
\providecommand \href@noop [0]{\@secondoftwo}%
\providecommand \href [0]{\begingroup \@sanitize@url \@href}%
\providecommand \@href[1]{\@@startlink{#1}\@@href}%
\providecommand \@@href[1]{\endgroup#1\@@endlink}%
\providecommand \@sanitize@url [0]{\catcode `\\12\catcode `\$12\catcode
  `\&12\catcode `\#12\catcode `\^12\catcode `\_12\catcode `\%12\relax}%
\providecommand \@@startlink[1]{}%
\providecommand \@@endlink[0]{}%
\providecommand \url  [0]{\begingroup\@sanitize@url \@url }%
\providecommand \@url [1]{\endgroup\@href {#1}{\urlprefix }}%
\providecommand \urlprefix  [0]{URL }%
\providecommand \Eprint [0]{\href }%
\providecommand \doibase [0]{http://dx.doi.org/}%
\providecommand \selectlanguage [0]{\@gobble}%
\providecommand \bibinfo  [0]{\@secondoftwo}%
\providecommand \bibfield  [0]{\@secondoftwo}%
\providecommand \translation [1]{[#1]}%
\providecommand \BibitemOpen [0]{}%
\providecommand \bibitemStop [0]{}%
\providecommand \bibitemNoStop [0]{.\EOS\space}%
\providecommand \EOS [0]{\spacefactor3000\relax}%
\providecommand \BibitemShut  [1]{\csname bibitem#1\endcsname}%
\let\auto@bib@innerbib\@empty
\bibitem [{\citenamefont {Cui}\ \emph {et~al.}(2017)\citenamefont {Cui} \emph
  {et~al.}}]{PandaX-II:2017hlx}%
  \BibitemOpen
  \bibfield  {author} {\bibinfo {author} {\bibfnamefont {X.}~\bibnamefont
  {Cui}} \emph {et~al.} (\bibinfo {collaboration} {PandaX-II}),\ }\href
  {\doibase 10.1103/PhysRevLett.119.181302} {\bibfield  {journal} {\bibinfo
  {journal} {Phys. Rev. Lett.}\ }\textbf {\bibinfo {volume} {119}},\ \bibinfo
  {pages} {181302} (\bibinfo {year} {2017})},\ \Eprint
  {http://arxiv.org/abs/1708.06917} {arXiv:1708.06917 [astro-ph.CO]}
  \BibitemShut {NoStop}%
\bibitem [{\citenamefont {Aprile}\ \emph {et~al.}(2018)\citenamefont {Aprile}
  \emph {et~al.}}]{XENON:2018voc}%
  \BibitemOpen
  \bibfield  {author} {\bibinfo {author} {\bibfnamefont {E.}~\bibnamefont
  {Aprile}} \emph {et~al.} (\bibinfo {collaboration} {XENON}),\ }\href
  {\doibase 10.1103/PhysRevLett.121.111302} {\bibfield  {journal} {\bibinfo
  {journal} {Phys. Rev. Lett.}\ }\textbf {\bibinfo {volume} {121}},\ \bibinfo
  {pages} {111302} (\bibinfo {year} {2018})},\ \Eprint
  {http://arxiv.org/abs/1805.12562} {arXiv:1805.12562 [astro-ph.CO]}
  \BibitemShut {NoStop}%
\bibitem [{\citenamefont {Aalbers}\ \emph {et~al.}(2023)\citenamefont {Aalbers}
  \emph {et~al.}}]{LZ:2022lsv}%
  \BibitemOpen
  \bibfield  {author} {\bibinfo {author} {\bibfnamefont {J.}~\bibnamefont
  {Aalbers}} \emph {et~al.} (\bibinfo {collaboration} {LZ}),\ }\href {\doibase
  10.1103/PhysRevLett.131.041002} {\bibfield  {journal} {\bibinfo  {journal}
  {Phys. Rev. Lett.}\ }\textbf {\bibinfo {volume} {131}},\ \bibinfo {pages}
  {041002} (\bibinfo {year} {2023})},\ \Eprint
  {http://arxiv.org/abs/2207.03764} {arXiv:2207.03764 [hep-ex]} \BibitemShut
  {NoStop}%
\bibitem [{\citenamefont {Aalbers}\ \emph {et~al.}(2025)\citenamefont {Aalbers}
  \emph {et~al.}}]{LZ:2024zvo}%
  \BibitemOpen
  \bibfield  {author} {\bibinfo {author} {\bibfnamefont {J.}~\bibnamefont
  {Aalbers}} \emph {et~al.} (\bibinfo {collaboration} {LZ}),\ }\href {\doibase
  10.1103/4dyc-z8zf} {\bibfield  {journal} {\bibinfo  {journal} {Phys. Rev.
  Lett.}\ }\textbf {\bibinfo {volume} {135}},\ \bibinfo {pages} {011802}
  (\bibinfo {year} {2025})},\ \Eprint {http://arxiv.org/abs/2410.17036}
  {arXiv:2410.17036 [hep-ex]} \BibitemShut {NoStop}%
\bibitem [{\citenamefont {Barak}\ \emph {et~al.}(2020)\citenamefont {Barak}
  \emph {et~al.}}]{SENSEI:2020dpa}%
  \BibitemOpen
  \bibfield  {author} {\bibinfo {author} {\bibfnamefont {L.}~\bibnamefont
  {Barak}} \emph {et~al.} (\bibinfo {collaboration} {SENSEI}),\ }\href
  {\doibase 10.1103/PhysRevLett.125.171802} {\bibfield  {journal} {\bibinfo
  {journal} {Phys. Rev. Lett.}\ }\textbf {\bibinfo {volume} {125}},\ \bibinfo
  {pages} {171802} (\bibinfo {year} {2020})},\ \Eprint
  {http://arxiv.org/abs/2004.11378} {arXiv:2004.11378 [astro-ph.CO]}
  \BibitemShut {NoStop}%
\bibitem [{\citenamefont {Li}\ \emph {et~al.}(2023)\citenamefont {Li} \emph
  {et~al.}}]{PandaX:2022xqx}%
  \BibitemOpen
  \bibfield  {author} {\bibinfo {author} {\bibfnamefont {S.}~\bibnamefont {Li}}
  \emph {et~al.} (\bibinfo {collaboration} {PandaX}),\ }\href {\doibase
  10.1103/PhysRevLett.130.261001} {\bibfield  {journal} {\bibinfo  {journal}
  {Phys. Rev. Lett.}\ }\textbf {\bibinfo {volume} {130}},\ \bibinfo {pages}
  {261001} (\bibinfo {year} {2023})},\ \Eprint
  {http://arxiv.org/abs/2212.10067} {arXiv:2212.10067 [hep-ex]} \BibitemShut
  {NoStop}%
\bibitem [{\citenamefont {Agnes}\ \emph {et~al.}(2023)\citenamefont {Agnes}
  \emph {et~al.}}]{DarkSide:2022knj}%
  \BibitemOpen
  \bibfield  {author} {\bibinfo {author} {\bibfnamefont {P.}~\bibnamefont
  {Agnes}} \emph {et~al.} (\bibinfo {collaboration} {DarkSide}),\ }\href
  {\doibase 10.1103/PhysRevLett.130.101002} {\bibfield  {journal} {\bibinfo
  {journal} {Phys. Rev. Lett.}\ }\textbf {\bibinfo {volume} {130}},\ \bibinfo
  {pages} {101002} (\bibinfo {year} {2023})},\ \Eprint
  {http://arxiv.org/abs/2207.11968} {arXiv:2207.11968 [hep-ex]} \BibitemShut
  {NoStop}%
\bibitem [{\citenamefont {Arnquist}\ \emph {et~al.}(2023)\citenamefont
  {Arnquist} \emph {et~al.}}]{DAMIC-M:2023gxo}%
  \BibitemOpen
  \bibfield  {author} {\bibinfo {author} {\bibfnamefont {I.}~\bibnamefont
  {Arnquist}} \emph {et~al.} (\bibinfo {collaboration} {DAMIC-M}),\ }\href
  {\doibase 10.1103/PhysRevLett.130.171003} {\bibfield  {journal} {\bibinfo
  {journal} {Phys. Rev. Lett.}\ }\textbf {\bibinfo {volume} {130}},\ \bibinfo
  {pages} {171003} (\bibinfo {year} {2023})},\ \Eprint
  {http://arxiv.org/abs/2302.02372} {arXiv:2302.02372 [hep-ex]} \BibitemShut
  {NoStop}%
\bibitem [{\citenamefont {Arnquist}\ \emph {et~al.}(2024)\citenamefont
  {Arnquist} \emph {et~al.}}]{DAMIC-M:2023hgj}%
  \BibitemOpen
  \bibfield  {author} {\bibinfo {author} {\bibfnamefont {I.}~\bibnamefont
  {Arnquist}} \emph {et~al.} (\bibinfo {collaboration} {DAMIC-M}),\ }\href
  {\doibase 10.1103/PhysRevLett.132.101006} {\bibfield  {journal} {\bibinfo
  {journal} {Phys. Rev. Lett.}\ }\textbf {\bibinfo {volume} {132}},\ \bibinfo
  {pages} {101006} (\bibinfo {year} {2024})},\ \Eprint
  {http://arxiv.org/abs/2307.07251} {arXiv:2307.07251 [hep-ex]} \BibitemShut
  {NoStop}%
\bibitem [{\citenamefont {Aggarwal}\ \emph
  {et~al.}(2025{\natexlab{a}})\citenamefont {Aggarwal} \emph
  {et~al.}}]{DAMIC-M:2025luv}%
  \BibitemOpen
  \bibfield  {author} {\bibinfo {author} {\bibfnamefont {K.}~\bibnamefont
  {Aggarwal}} \emph {et~al.} (\bibinfo {collaboration} {DAMIC-M}),\ }\href
  {\doibase 10.1103/2tcc-bqck} {\bibfield  {journal} {\bibinfo  {journal}
  {Phys. Rev. Lett.}\ }\textbf {\bibinfo {volume} {135}},\ \bibinfo {pages}
  {071002} (\bibinfo {year} {2025}{\natexlab{a}})},\ \Eprint
  {http://arxiv.org/abs/2503.14617} {arXiv:2503.14617 [hep-ex]} \BibitemShut
  {NoStop}%
\bibitem [{\citenamefont {Zhang}\ \emph {et~al.}(2025)\citenamefont {Zhang}
  \emph {et~al.}}]{Zhang:2025ajc}%
  \BibitemOpen
  \bibfield  {author} {\bibinfo {author} {\bibfnamefont {M.}~\bibnamefont
  {Zhang}} \emph {et~al.},\ }\href@noop {} {\  (\bibinfo {year} {2025})},\
  \Eprint {http://arxiv.org/abs/2507.11930} {arXiv:2507.11930 [hep-ex]}
  \BibitemShut {NoStop}%
\bibitem [{\citenamefont {Hall}\ \emph {et~al.}(2010)\citenamefont {Hall},
  \citenamefont {Jedamzik}, \citenamefont {March-Russell},\ and\ \citenamefont
  {West}}]{Hall:2009bx}%
  \BibitemOpen
  \bibfield  {author} {\bibinfo {author} {\bibfnamefont {L.~J.}\ \bibnamefont
  {Hall}}, \bibinfo {author} {\bibfnamefont {K.}~\bibnamefont {Jedamzik}},
  \bibinfo {author} {\bibfnamefont {J.}~\bibnamefont {March-Russell}}, \ and\
  \bibinfo {author} {\bibfnamefont {S.~M.}\ \bibnamefont {West}},\ }\href
  {\doibase 10.1007/JHEP03(2010)080} {\bibfield  {journal} {\bibinfo  {journal}
  {JHEP}\ }\textbf {\bibinfo {volume} {03}},\ \bibinfo {pages} {080} (\bibinfo
  {year} {2010})},\ \Eprint {http://arxiv.org/abs/0911.1120} {arXiv:0911.1120
  [hep-ph]} \BibitemShut {NoStop}%
\bibitem [{\citenamefont {Chu}\ \emph {et~al.}(2012)\citenamefont {Chu},
  \citenamefont {Hambye},\ and\ \citenamefont {Tytgat}}]{Chu:2011be}%
  \BibitemOpen
  \bibfield  {author} {\bibinfo {author} {\bibfnamefont {X.}~\bibnamefont
  {Chu}}, \bibinfo {author} {\bibfnamefont {T.}~\bibnamefont {Hambye}}, \ and\
  \bibinfo {author} {\bibfnamefont {M.~H.~G.}\ \bibnamefont {Tytgat}},\ }\href
  {\doibase 10.1088/1475-7516/2012/05/034} {\bibfield  {journal} {\bibinfo
  {journal} {JCAP}\ }\textbf {\bibinfo {volume} {05}},\ \bibinfo {pages} {034}
  (\bibinfo {year} {2012})},\ \Eprint {http://arxiv.org/abs/1112.0493}
  {arXiv:1112.0493 [hep-ph]} \BibitemShut {NoStop}%
\bibitem [{\citenamefont {Essig}\ \emph {et~al.}(2012)\citenamefont {Essig},
  \citenamefont {Mardon},\ and\ \citenamefont {Volansky}}]{Essig:2011nj}%
  \BibitemOpen
  \bibfield  {author} {\bibinfo {author} {\bibfnamefont {R.}~\bibnamefont
  {Essig}}, \bibinfo {author} {\bibfnamefont {J.}~\bibnamefont {Mardon}}, \
  and\ \bibinfo {author} {\bibfnamefont {T.}~\bibnamefont {Volansky}},\ }\href
  {\doibase 10.1103/PhysRevD.85.076007} {\bibfield  {journal} {\bibinfo
  {journal} {Phys. Rev. D}\ }\textbf {\bibinfo {volume} {85}},\ \bibinfo
  {pages} {076007} (\bibinfo {year} {2012})},\ \Eprint
  {http://arxiv.org/abs/1108.5383} {arXiv:1108.5383 [hep-ph]} \BibitemShut
  {NoStop}%
\bibitem [{\citenamefont {Boddy}\ \emph {et~al.}(2025)\citenamefont {Boddy},
  \citenamefont {Freese}, \citenamefont {Montefalcone},\ and\ \citenamefont
  {Shams Es~Haghi}}]{Boddy:2024vgt}%
  \BibitemOpen
  \bibfield  {author} {\bibinfo {author} {\bibfnamefont {K.~K.}\ \bibnamefont
  {Boddy}}, \bibinfo {author} {\bibfnamefont {K.}~\bibnamefont {Freese}},
  \bibinfo {author} {\bibfnamefont {G.}~\bibnamefont {Montefalcone}}, \ and\
  \bibinfo {author} {\bibfnamefont {B.}~\bibnamefont {Shams Es~Haghi}},\ }\href
  {\doibase 10.1103/PhysRevD.111.063537} {\bibfield  {journal} {\bibinfo
  {journal} {Phys. Rev. D}\ }\textbf {\bibinfo {volume} {111}},\ \bibinfo
  {pages} {063537} (\bibinfo {year} {2025})},\ \Eprint
  {http://arxiv.org/abs/2405.06226} {arXiv:2405.06226 [hep-ph]} \BibitemShut
  {NoStop}%
\bibitem [{\citenamefont {Bernal}\ \emph {et~al.}(2025)\citenamefont {Bernal},
  \citenamefont {Fong},\ and\ \citenamefont {Zapata}}]{Bernal:2024ndy}%
  \BibitemOpen
  \bibfield  {author} {\bibinfo {author} {\bibfnamefont {N.}~\bibnamefont
  {Bernal}}, \bibinfo {author} {\bibfnamefont {C.~S.}\ \bibnamefont {Fong}}, \
  and\ \bibinfo {author} {\bibfnamefont {{\'O}.}~\bibnamefont {Zapata}},\
  }\href {\doibase 10.1007/JHEP02(2025)161} {\bibfield  {journal} {\bibinfo
  {journal} {JHEP}\ }\textbf {\bibinfo {volume} {02}},\ \bibinfo {pages} {161}
  (\bibinfo {year} {2025})},\ \Eprint {http://arxiv.org/abs/2412.04550}
  {arXiv:2412.04550 [hep-ph]} \BibitemShut {NoStop}%
\bibitem [{\citenamefont {Garny}\ \emph {et~al.}(2016)\citenamefont {Garny},
  \citenamefont {Sandora},\ and\ \citenamefont {Sloth}}]{Garny:2015sjg}%
  \BibitemOpen
  \bibfield  {author} {\bibinfo {author} {\bibfnamefont {M.}~\bibnamefont
  {Garny}}, \bibinfo {author} {\bibfnamefont {M.}~\bibnamefont {Sandora}}, \
  and\ \bibinfo {author} {\bibfnamefont {M.~S.}\ \bibnamefont {Sloth}},\ }\href
  {\doibase 10.1103/PhysRevLett.116.101302} {\bibfield  {journal} {\bibinfo
  {journal} {Phys. Rev. Lett.}\ }\textbf {\bibinfo {volume} {116}},\ \bibinfo
  {pages} {101302} (\bibinfo {year} {2016})},\ \Eprint
  {http://arxiv.org/abs/1511.03278} {arXiv:1511.03278 [hep-ph]} \BibitemShut
  {NoStop}%
\bibitem [{\citenamefont {Tang}\ and\ \citenamefont {Wu}(2017)}]{Tang:2017hvq}%
  \BibitemOpen
  \bibfield  {author} {\bibinfo {author} {\bibfnamefont {Y.}~\bibnamefont
  {Tang}}\ and\ \bibinfo {author} {\bibfnamefont {Y.-L.}\ \bibnamefont {Wu}},\
  }\href {\doibase 10.1016/j.physletb.2017.10.034} {\bibfield  {journal}
  {\bibinfo  {journal} {Phys. Lett. B}\ }\textbf {\bibinfo {volume} {774}},\
  \bibinfo {pages} {676} (\bibinfo {year} {2017})},\ \Eprint
  {http://arxiv.org/abs/1708.05138} {arXiv:1708.05138 [hep-ph]} \BibitemShut
  {NoStop}%
\bibitem [{\citenamefont {Garny}\ \emph {et~al.}(2018)\citenamefont {Garny},
  \citenamefont {Palessandro}, \citenamefont {Sandora},\ and\ \citenamefont
  {Sloth}}]{Garny:2017kha}%
  \BibitemOpen
  \bibfield  {author} {\bibinfo {author} {\bibfnamefont {M.}~\bibnamefont
  {Garny}}, \bibinfo {author} {\bibfnamefont {A.}~\bibnamefont {Palessandro}},
  \bibinfo {author} {\bibfnamefont {M.}~\bibnamefont {Sandora}}, \ and\
  \bibinfo {author} {\bibfnamefont {M.~S.}\ \bibnamefont {Sloth}},\ }\href
  {\doibase 10.1088/1475-7516/2018/02/027} {\bibfield  {journal} {\bibinfo
  {journal} {JCAP}\ }\textbf {\bibinfo {volume} {02}},\ \bibinfo {pages} {027}
  (\bibinfo {year} {2018})},\ \Eprint {http://arxiv.org/abs/1709.09688}
  {arXiv:1709.09688 [hep-ph]} \BibitemShut {NoStop}%
\bibitem [{\citenamefont {Bernal}\ \emph {et~al.}(2018)\citenamefont {Bernal},
  \citenamefont {Dutra}, \citenamefont {Mambrini}, \citenamefont {Olive},
  \citenamefont {Peloso},\ and\ \citenamefont {Pierre}}]{Bernal:2018qlk}%
  \BibitemOpen
  \bibfield  {author} {\bibinfo {author} {\bibfnamefont {N.}~\bibnamefont
  {Bernal}}, \bibinfo {author} {\bibfnamefont {M.}~\bibnamefont {Dutra}},
  \bibinfo {author} {\bibfnamefont {Y.}~\bibnamefont {Mambrini}}, \bibinfo
  {author} {\bibfnamefont {K.}~\bibnamefont {Olive}}, \bibinfo {author}
  {\bibfnamefont {M.}~\bibnamefont {Peloso}}, \ and\ \bibinfo {author}
  {\bibfnamefont {M.}~\bibnamefont {Pierre}},\ }\href {\doibase
  10.1103/PhysRevD.97.115020} {\bibfield  {journal} {\bibinfo  {journal} {Phys.
  Rev. D}\ }\textbf {\bibinfo {volume} {97}},\ \bibinfo {pages} {115020}
  (\bibinfo {year} {2018})},\ \Eprint {http://arxiv.org/abs/1803.01866}
  {arXiv:1803.01866 [hep-ph]} \BibitemShut {NoStop}%
\bibitem [{\citenamefont {Elahi}\ \emph {et~al.}(2015)\citenamefont {Elahi},
  \citenamefont {Kolda},\ and\ \citenamefont {Unwin}}]{Elahi:2014fsa}%
  \BibitemOpen
  \bibfield  {author} {\bibinfo {author} {\bibfnamefont {F.}~\bibnamefont
  {Elahi}}, \bibinfo {author} {\bibfnamefont {C.}~\bibnamefont {Kolda}}, \ and\
  \bibinfo {author} {\bibfnamefont {J.}~\bibnamefont {Unwin}},\ }\href
  {\doibase 10.1007/JHEP03(2015)048} {\bibfield  {journal} {\bibinfo  {journal}
  {JHEP}\ }\textbf {\bibinfo {volume} {03}},\ \bibinfo {pages} {048} (\bibinfo
  {year} {2015})},\ \Eprint {http://arxiv.org/abs/1410.6157} {arXiv:1410.6157
  [hep-ph]} \BibitemShut {NoStop}%
\bibitem [{\citenamefont {Baryakhtar}\ \emph {et~al.}(2017)\citenamefont
  {Baryakhtar}, \citenamefont {Lasenby},\ and\ \citenamefont
  {Teo}}]{Baryakhtar:2017ngi}%
  \BibitemOpen
  \bibfield  {author} {\bibinfo {author} {\bibfnamefont {M.}~\bibnamefont
  {Baryakhtar}}, \bibinfo {author} {\bibfnamefont {R.}~\bibnamefont {Lasenby}},
  \ and\ \bibinfo {author} {\bibfnamefont {M.}~\bibnamefont {Teo}},\ }\href
  {\doibase 10.1103/PhysRevD.96.035019} {\bibfield  {journal} {\bibinfo
  {journal} {Phys. Rev. D}\ }\textbf {\bibinfo {volume} {96}},\ \bibinfo
  {pages} {035019} (\bibinfo {year} {2017})},\ \Eprint
  {http://arxiv.org/abs/1704.05081} {arXiv:1704.05081 [hep-ph]} \BibitemShut
  {NoStop}%
\bibitem [{\citenamefont {Siemonsen}\ \emph {et~al.}(2023)\citenamefont
  {Siemonsen}, \citenamefont {Mondino}, \citenamefont {Egana-Ugrinovic},
  \citenamefont {Huang}, \citenamefont {Baryakhtar},\ and\ \citenamefont
  {East}}]{Siemonsen:2022ivj}%
  \BibitemOpen
  \bibfield  {author} {\bibinfo {author} {\bibfnamefont {N.}~\bibnamefont
  {Siemonsen}}, \bibinfo {author} {\bibfnamefont {C.}~\bibnamefont {Mondino}},
  \bibinfo {author} {\bibfnamefont {D.}~\bibnamefont {Egana-Ugrinovic}},
  \bibinfo {author} {\bibfnamefont {J.}~\bibnamefont {Huang}}, \bibinfo
  {author} {\bibfnamefont {M.}~\bibnamefont {Baryakhtar}}, \ and\ \bibinfo
  {author} {\bibfnamefont {W.~E.}\ \bibnamefont {East}},\ }\href {\doibase
  10.1103/PhysRevD.107.075025} {\bibfield  {journal} {\bibinfo  {journal}
  {Phys. Rev. D}\ }\textbf {\bibinfo {volume} {107}},\ \bibinfo {pages}
  {075025} (\bibinfo {year} {2023})},\ \Eprint
  {http://arxiv.org/abs/2212.09772} {arXiv:2212.09772 [astro-ph.HE]}
  \BibitemShut {NoStop}%
\bibitem [{\citenamefont {Fixsen}\ \emph {et~al.}(1996)\citenamefont {Fixsen},
  \citenamefont {Cheng}, \citenamefont {Gales}, \citenamefont {Mather},
  \citenamefont {Shafer},\ and\ \citenamefont {Wright}}]{Fixsen:1996nj}%
  \BibitemOpen
  \bibfield  {author} {\bibinfo {author} {\bibfnamefont {D.~J.}\ \bibnamefont
  {Fixsen}}, \bibinfo {author} {\bibfnamefont {E.~S.}\ \bibnamefont {Cheng}},
  \bibinfo {author} {\bibfnamefont {J.~M.}\ \bibnamefont {Gales}}, \bibinfo
  {author} {\bibfnamefont {J.~C.}\ \bibnamefont {Mather}}, \bibinfo {author}
  {\bibfnamefont {R.~A.}\ \bibnamefont {Shafer}}, \ and\ \bibinfo {author}
  {\bibfnamefont {E.~L.}\ \bibnamefont {Wright}},\ }\href {\doibase
  10.1086/178173} {\bibfield  {journal} {\bibinfo  {journal} {Astrophys. J.}\
  }\textbf {\bibinfo {volume} {473}},\ \bibinfo {pages} {576} (\bibinfo {year}
  {1996})},\ \Eprint {http://arxiv.org/abs/astro-ph/9605054}
  {arXiv:astro-ph/9605054} \BibitemShut {NoStop}%
\bibitem [{\citenamefont {Caputo}\ \emph {et~al.}(2020)\citenamefont {Caputo},
  \citenamefont {Liu}, \citenamefont {Mishra-Sharma},\ and\ \citenamefont
  {Ruderman}}]{Caputo:2020bdy}%
  \BibitemOpen
  \bibfield  {author} {\bibinfo {author} {\bibfnamefont {A.}~\bibnamefont
  {Caputo}}, \bibinfo {author} {\bibfnamefont {H.}~\bibnamefont {Liu}},
  \bibinfo {author} {\bibfnamefont {S.}~\bibnamefont {Mishra-Sharma}}, \ and\
  \bibinfo {author} {\bibfnamefont {J.~T.}\ \bibnamefont {Ruderman}},\ }\href
  {\doibase 10.1103/PhysRevLett.125.221303} {\bibfield  {journal} {\bibinfo
  {journal} {Phys. Rev. Lett.}\ }\textbf {\bibinfo {volume} {125}},\ \bibinfo
  {pages} {221303} (\bibinfo {year} {2020})},\ \Eprint
  {http://arxiv.org/abs/2002.05165} {arXiv:2002.05165 [astro-ph.CO]}
  \BibitemShut {NoStop}%
\bibitem [{\citenamefont {Bhattiprolu}\ \emph {et~al.}(2024)\citenamefont
  {Bhattiprolu}, \citenamefont {McGehee},\ and\ \citenamefont
  {Pierce}}]{Bhattiprolu:2023akk}%
  \BibitemOpen
  \bibfield  {author} {\bibinfo {author} {\bibfnamefont {P.~N.}\ \bibnamefont
  {Bhattiprolu}}, \bibinfo {author} {\bibfnamefont {R.}~\bibnamefont
  {McGehee}}, \ and\ \bibinfo {author} {\bibfnamefont {A.}~\bibnamefont
  {Pierce}},\ }\href {\doibase 10.1103/PhysRevD.110.L031702} {\bibfield
  {journal} {\bibinfo  {journal} {Phys. Rev. D}\ }\textbf {\bibinfo {volume}
  {110}},\ \bibinfo {pages} {L031702} (\bibinfo {year} {2024})},\ \Eprint
  {http://arxiv.org/abs/2312.14152} {arXiv:2312.14152 [hep-ph]} \BibitemShut
  {NoStop}%
\bibitem [{\citenamefont {Fernandez}\ \emph {et~al.}(2022)\citenamefont
  {Fernandez}, \citenamefont {Kahn},\ and\ \citenamefont
  {Shelton}}]{Fernandez:2021iti}%
  \BibitemOpen
  \bibfield  {author} {\bibinfo {author} {\bibfnamefont {N.}~\bibnamefont
  {Fernandez}}, \bibinfo {author} {\bibfnamefont {Y.}~\bibnamefont {Kahn}}, \
  and\ \bibinfo {author} {\bibfnamefont {J.}~\bibnamefont {Shelton}},\ }\href
  {\doibase 10.1007/JHEP07(2022)044} {\bibfield  {journal} {\bibinfo  {journal}
  {JHEP}\ }\textbf {\bibinfo {volume} {07}},\ \bibinfo {pages} {044} (\bibinfo
  {year} {2022})},\ \Eprint {http://arxiv.org/abs/2111.13709} {arXiv:2111.13709
  [hep-ph]} \BibitemShut {NoStop}%
\bibitem [{\citenamefont {Hambye}\ \emph {et~al.}(2018)\citenamefont {Hambye},
  \citenamefont {Tytgat}, \citenamefont {Vandecasteele},\ and\ \citenamefont
  {Vanderheyden}}]{Hambye:2018dpi}%
  \BibitemOpen
  \bibfield  {author} {\bibinfo {author} {\bibfnamefont {T.}~\bibnamefont
  {Hambye}}, \bibinfo {author} {\bibfnamefont {M.~H.~G.}\ \bibnamefont
  {Tytgat}}, \bibinfo {author} {\bibfnamefont {J.}~\bibnamefont
  {Vandecasteele}}, \ and\ \bibinfo {author} {\bibfnamefont {L.}~\bibnamefont
  {Vanderheyden}},\ }\href {\doibase 10.1103/PhysRevD.98.075017} {\bibfield
  {journal} {\bibinfo  {journal} {Phys. Rev. D}\ }\textbf {\bibinfo {volume}
  {98}},\ \bibinfo {pages} {075017} (\bibinfo {year} {2018})},\ \Eprint
  {http://arxiv.org/abs/1807.05022} {arXiv:1807.05022 [hep-ph]} \BibitemShut
  {NoStop}%
\bibitem [{\citenamefont {Leane}\ and\ \citenamefont
  {Beacom}(2025)}]{Leane:2025efj}%
  \BibitemOpen
  \bibfield  {author} {\bibinfo {author} {\bibfnamefont {R.~K.}\ \bibnamefont
  {Leane}}\ and\ \bibinfo {author} {\bibfnamefont {J.~F.}\ \bibnamefont
  {Beacom}},\ }\href {\doibase 10.1103/mcyx-g4pd} {\  (\bibinfo {year}
  {2025}),\ 10.1103/mcyx-g4pd},\ \Eprint {http://arxiv.org/abs/2503.09685}
  {arXiv:2503.09685 [hep-ph]} \BibitemShut {NoStop}%
\bibitem [{\citenamefont {Hasegawa}\ \emph {et~al.}(2019)\citenamefont
  {Hasegawa}, \citenamefont {Hiroshima}, \citenamefont {Kohri}, \citenamefont
  {Hansen}, \citenamefont {Tram},\ and\ \citenamefont
  {Hannestad}}]{Hasegawa:2019jsa}%
  \BibitemOpen
  \bibfield  {author} {\bibinfo {author} {\bibfnamefont {T.}~\bibnamefont
  {Hasegawa}}, \bibinfo {author} {\bibfnamefont {N.}~\bibnamefont {Hiroshima}},
  \bibinfo {author} {\bibfnamefont {K.}~\bibnamefont {Kohri}}, \bibinfo
  {author} {\bibfnamefont {R.~S.~L.}\ \bibnamefont {Hansen}}, \bibinfo {author}
  {\bibfnamefont {T.}~\bibnamefont {Tram}}, \ and\ \bibinfo {author}
  {\bibfnamefont {S.}~\bibnamefont {Hannestad}},\ }\href {\doibase
  10.1088/1475-7516/2019/12/012} {\bibfield  {journal} {\bibinfo  {journal}
  {JCAP}\ }\textbf {\bibinfo {volume} {12}},\ \bibinfo {pages} {012} (\bibinfo
  {year} {2019})},\ \Eprint {http://arxiv.org/abs/1908.10189} {arXiv:1908.10189
  [hep-ph]} \BibitemShut {NoStop}%
\bibitem [{\citenamefont {Kawasaki}\ \emph {et~al.}(2000)\citenamefont
  {Kawasaki}, \citenamefont {Kohri},\ and\ \citenamefont
  {Sugiyama}}]{Kawasaki:2000en}%
  \BibitemOpen
  \bibfield  {author} {\bibinfo {author} {\bibfnamefont {M.}~\bibnamefont
  {Kawasaki}}, \bibinfo {author} {\bibfnamefont {K.}~\bibnamefont {Kohri}}, \
  and\ \bibinfo {author} {\bibfnamefont {N.}~\bibnamefont {Sugiyama}},\ }\href
  {\doibase 10.1103/PhysRevD.62.023506} {\bibfield  {journal} {\bibinfo
  {journal} {Phys. Rev. D}\ }\textbf {\bibinfo {volume} {62}},\ \bibinfo
  {pages} {023506} (\bibinfo {year} {2000})},\ \Eprint
  {http://arxiv.org/abs/astro-ph/0002127} {arXiv:astro-ph/0002127} \BibitemShut
  {NoStop}%
\bibitem [{\citenamefont {{Zeldovich}}\ and\ \citenamefont
  {{Khlopov}}(1978)}]{1978PhLB...79..239Z}%
  \BibitemOpen
  \bibfield  {author} {\bibinfo {author} {\bibfnamefont {Y.~B.}\ \bibnamefont
  {{Zeldovich}}}\ and\ \bibinfo {author} {\bibfnamefont {M.~Y.}\ \bibnamefont
  {{Khlopov}}},\ }\href {\doibase 10.1016/0370-2693(78)90232-0} {\bibfield
  {journal} {\bibinfo  {journal} {Physics Letters B}\ }\textbf {\bibinfo
  {volume} {79}},\ \bibinfo {pages} {239} (\bibinfo {year} {1978})}\BibitemShut
  {NoStop}%
\bibitem [{\citenamefont {Preskill}(1979)}]{Preskill:1979zi}%
  \BibitemOpen
  \bibfield  {author} {\bibinfo {author} {\bibfnamefont {J.}~\bibnamefont
  {Preskill}},\ }\href {\doibase 10.1103/PhysRevLett.43.1365} {\bibfield
  {journal} {\bibinfo  {journal} {Phys. Rev. Lett.}\ }\textbf {\bibinfo
  {volume} {43}},\ \bibinfo {pages} {1365} (\bibinfo {year}
  {1979})}\BibitemShut {NoStop}%
\bibitem [{\citenamefont {{Moroi}}\ \emph {et~al.}(1993)\citenamefont
  {{Moroi}}, \citenamefont {{Murayama}},\ and\ \citenamefont
  {{Yamaguchi}}}]{1993PhLB..303..289M}%
  \BibitemOpen
  \bibfield  {author} {\bibinfo {author} {\bibfnamefont {T.}~\bibnamefont
  {{Moroi}}}, \bibinfo {author} {\bibfnamefont {H.}~\bibnamefont {{Murayama}}},
  \ and\ \bibinfo {author} {\bibfnamefont {M.}~\bibnamefont {{Yamaguchi}}},\
  }\href {\doibase 10.1016/0370-2693(93)91434-O} {\bibfield  {journal}
  {\bibinfo  {journal} {Physics Letters B}\ }\textbf {\bibinfo {volume}
  {303}},\ \bibinfo {pages} {289} (\bibinfo {year} {1993})}\BibitemShut
  {NoStop}%
\bibitem [{\citenamefont {Asaka}\ \emph {et~al.}(2000)\citenamefont {Asaka},
  \citenamefont {Hamaguchi},\ and\ \citenamefont {Suzuki}}]{Asaka:2000zh}%
  \BibitemOpen
  \bibfield  {author} {\bibinfo {author} {\bibfnamefont {T.}~\bibnamefont
  {Asaka}}, \bibinfo {author} {\bibfnamefont {K.}~\bibnamefont {Hamaguchi}}, \
  and\ \bibinfo {author} {\bibfnamefont {K.}~\bibnamefont {Suzuki}},\ }\href
  {\doibase 10.1016/S0370-2693(00)00959-X} {\bibfield  {journal} {\bibinfo
  {journal} {Phys. Lett. B}\ }\textbf {\bibinfo {volume} {490}},\ \bibinfo
  {pages} {136} (\bibinfo {year} {2000})},\ \Eprint
  {http://arxiv.org/abs/hep-ph/0005136} {arXiv:hep-ph/0005136} \BibitemShut
  {NoStop}%
\bibitem [{\citenamefont {Pradler}\ and\ \citenamefont
  {Steffen}(2007)}]{Pradler:2006hh}%
  \BibitemOpen
  \bibfield  {author} {\bibinfo {author} {\bibfnamefont {J.}~\bibnamefont
  {Pradler}}\ and\ \bibinfo {author} {\bibfnamefont {F.~D.}\ \bibnamefont
  {Steffen}},\ }\href {\doibase 10.1016/j.physletb.2007.02.072} {\bibfield
  {journal} {\bibinfo  {journal} {Phys. Lett. B}\ }\textbf {\bibinfo {volume}
  {648}},\ \bibinfo {pages} {224} (\bibinfo {year} {2007})},\ \Eprint
  {http://arxiv.org/abs/hep-ph/0612291} {arXiv:hep-ph/0612291} \BibitemShut
  {NoStop}%
\bibitem [{\citenamefont {Steffen}(2008)}]{Steffen:2008bt}%
  \BibitemOpen
  \bibfield  {author} {\bibinfo {author} {\bibfnamefont {F.~D.}\ \bibnamefont
  {Steffen}},\ }\href {\doibase 10.1016/j.physletb.2008.09.036} {\bibfield
  {journal} {\bibinfo  {journal} {Phys. Lett. B}\ }\textbf {\bibinfo {volume}
  {669}},\ \bibinfo {pages} {74} (\bibinfo {year} {2008})},\ \Eprint
  {http://arxiv.org/abs/0806.3266} {arXiv:0806.3266 [hep-ph]} \BibitemShut
  {NoStop}%
\bibitem [{\citenamefont {Covi}\ \emph {et~al.}(2011)\citenamefont {Covi},
  \citenamefont {Olechowski}, \citenamefont {Pokorski}, \citenamefont
  {Turzynski},\ and\ \citenamefont {Wells}}]{Covi:2010au}%
  \BibitemOpen
  \bibfield  {author} {\bibinfo {author} {\bibfnamefont {L.}~\bibnamefont
  {Covi}}, \bibinfo {author} {\bibfnamefont {M.}~\bibnamefont {Olechowski}},
  \bibinfo {author} {\bibfnamefont {S.}~\bibnamefont {Pokorski}}, \bibinfo
  {author} {\bibfnamefont {K.}~\bibnamefont {Turzynski}}, \ and\ \bibinfo
  {author} {\bibfnamefont {J.~D.}\ \bibnamefont {Wells}},\ }\href {\doibase
  10.1007/JHEP01(2011)033} {\bibfield  {journal} {\bibinfo  {journal} {JHEP}\
  }\textbf {\bibinfo {volume} {01}},\ \bibinfo {pages} {033} (\bibinfo {year}
  {2011})},\ \Eprint {http://arxiv.org/abs/1009.3801} {arXiv:1009.3801
  [hep-ph]} \BibitemShut {NoStop}%
\bibitem [{\citenamefont {Heisig}(2014)}]{Heisig:2013sva}%
  \BibitemOpen
  \bibfield  {author} {\bibinfo {author} {\bibfnamefont {J.}~\bibnamefont
  {Heisig}},\ }\href {\doibase 10.1088/1475-7516/2014/04/023} {\bibfield
  {journal} {\bibinfo  {journal} {JCAP}\ }\textbf {\bibinfo {volume} {04}},\
  \bibinfo {pages} {023} (\bibinfo {year} {2014})},\ \Eprint
  {http://arxiv.org/abs/1310.6352} {arXiv:1310.6352 [hep-ph]} \BibitemShut
  {NoStop}%
\bibitem [{\citenamefont {Starobinsky}(1980)}]{Starobinsky:1980te}%
  \BibitemOpen
  \bibfield  {author} {\bibinfo {author} {\bibfnamefont {A.~A.}\ \bibnamefont
  {Starobinsky}},\ }\href {\doibase 10.1016/0370-2693(80)90670-X} {\bibfield
  {journal} {\bibinfo  {journal} {Phys. Lett. B}\ }\textbf {\bibinfo {volume}
  {91}},\ \bibinfo {pages} {99} (\bibinfo {year} {1980})}\BibitemShut {NoStop}%
\bibitem [{\citenamefont {Guth}(1981)}]{Guth:1980zm}%
  \BibitemOpen
  \bibfield  {author} {\bibinfo {author} {\bibfnamefont {A.~H.}\ \bibnamefont
  {Guth}},\ }\href {\doibase 10.1103/PhysRevD.23.347} {\bibfield  {journal}
  {\bibinfo  {journal} {Phys. Rev. D}\ }\textbf {\bibinfo {volume} {23}},\
  \bibinfo {pages} {347} (\bibinfo {year} {1981})}\BibitemShut {NoStop}%
\bibitem [{\citenamefont {Linde}(1982)}]{Linde:1981mu}%
  \BibitemOpen
  \bibfield  {author} {\bibinfo {author} {\bibfnamefont {A.~D.}\ \bibnamefont
  {Linde}},\ }\href {\doibase 10.1016/0370-2693(82)91219-9} {\bibfield
  {journal} {\bibinfo  {journal} {Phys. Lett. B}\ }\textbf {\bibinfo {volume}
  {108}},\ \bibinfo {pages} {389} (\bibinfo {year} {1982})}\BibitemShut
  {NoStop}%
\bibitem [{\citenamefont {Albrecht}\ and\ \citenamefont
  {Steinhardt}(1982)}]{Albrecht:1982wi}%
  \BibitemOpen
  \bibfield  {author} {\bibinfo {author} {\bibfnamefont {A.}~\bibnamefont
  {Albrecht}}\ and\ \bibinfo {author} {\bibfnamefont {P.~J.}\ \bibnamefont
  {Steinhardt}},\ }\href {\doibase 10.1103/PhysRevLett.48.1220} {\bibfield
  {journal} {\bibinfo  {journal} {Phys. Rev. Lett.}\ }\textbf {\bibinfo
  {volume} {48}},\ \bibinfo {pages} {1220} (\bibinfo {year}
  {1982})}\BibitemShut {NoStop}%
\bibitem [{\citenamefont {Akrami}\ \emph {et~al.}(2020)\citenamefont {Akrami}
  \emph {et~al.}}]{Planck:2018jri}%
  \BibitemOpen
  \bibfield  {author} {\bibinfo {author} {\bibfnamefont {Y.}~\bibnamefont
  {Akrami}} \emph {et~al.} (\bibinfo {collaboration} {Planck}),\ }\href
  {\doibase 10.1051/0004-6361/201833887} {\bibfield  {journal} {\bibinfo
  {journal} {Astron. Astrophys.}\ }\textbf {\bibinfo {volume} {641}},\ \bibinfo
  {pages} {A10} (\bibinfo {year} {2020})},\ \Eprint
  {http://arxiv.org/abs/1807.06211} {arXiv:1807.06211 [astro-ph.CO]}
  \BibitemShut {NoStop}%
\bibitem [{\citenamefont {Khoury}\ \emph {et~al.}(2001)\citenamefont {Khoury},
  \citenamefont {Ovrut}, \citenamefont {Steinhardt},\ and\ \citenamefont
  {Turok}}]{Khoury:2001wf}%
  \BibitemOpen
  \bibfield  {author} {\bibinfo {author} {\bibfnamefont {J.}~\bibnamefont
  {Khoury}}, \bibinfo {author} {\bibfnamefont {B.~A.}\ \bibnamefont {Ovrut}},
  \bibinfo {author} {\bibfnamefont {P.~J.}\ \bibnamefont {Steinhardt}}, \ and\
  \bibinfo {author} {\bibfnamefont {N.}~\bibnamefont {Turok}},\ }\href
  {\doibase 10.1103/PhysRevD.64.123522} {\bibfield  {journal} {\bibinfo
  {journal} {Phys. Rev. D}\ }\textbf {\bibinfo {volume} {64}},\ \bibinfo
  {pages} {123522} (\bibinfo {year} {2001})},\ \Eprint
  {http://arxiv.org/abs/hep-th/0103239} {arXiv:hep-th/0103239} \BibitemShut
  {NoStop}%
\bibitem [{\citenamefont {Khoury}\ \emph
  {et~al.}(2002{\natexlab{a}})\citenamefont {Khoury}, \citenamefont {Ovrut},
  \citenamefont {Seiberg}, \citenamefont {Steinhardt},\ and\ \citenamefont
  {Turok}}]{Khoury:2001bz}%
  \BibitemOpen
  \bibfield  {author} {\bibinfo {author} {\bibfnamefont {J.}~\bibnamefont
  {Khoury}}, \bibinfo {author} {\bibfnamefont {B.~A.}\ \bibnamefont {Ovrut}},
  \bibinfo {author} {\bibfnamefont {N.}~\bibnamefont {Seiberg}}, \bibinfo
  {author} {\bibfnamefont {P.~J.}\ \bibnamefont {Steinhardt}}, \ and\ \bibinfo
  {author} {\bibfnamefont {N.}~\bibnamefont {Turok}},\ }\href {\doibase
  10.1103/PhysRevD.65.086007} {\bibfield  {journal} {\bibinfo  {journal} {Phys.
  Rev. D}\ }\textbf {\bibinfo {volume} {65}},\ \bibinfo {pages} {086007}
  (\bibinfo {year} {2002}{\natexlab{a}})},\ \Eprint
  {http://arxiv.org/abs/hep-th/0108187} {arXiv:hep-th/0108187} \BibitemShut
  {NoStop}%
\bibitem [{\citenamefont {Khoury}\ \emph
  {et~al.}(2002{\natexlab{b}})\citenamefont {Khoury}, \citenamefont {Ovrut},
  \citenamefont {Steinhardt},\ and\ \citenamefont {Turok}}]{Khoury:2001zk}%
  \BibitemOpen
  \bibfield  {author} {\bibinfo {author} {\bibfnamefont {J.}~\bibnamefont
  {Khoury}}, \bibinfo {author} {\bibfnamefont {B.~A.}\ \bibnamefont {Ovrut}},
  \bibinfo {author} {\bibfnamefont {P.~J.}\ \bibnamefont {Steinhardt}}, \ and\
  \bibinfo {author} {\bibfnamefont {N.}~\bibnamefont {Turok}},\ }\href
  {\doibase 10.1103/PhysRevD.66.046005} {\bibfield  {journal} {\bibinfo
  {journal} {Phys. Rev. D}\ }\textbf {\bibinfo {volume} {66}},\ \bibinfo
  {pages} {046005} (\bibinfo {year} {2002}{\natexlab{b}})},\ \Eprint
  {http://arxiv.org/abs/hep-th/0109050} {arXiv:hep-th/0109050} \BibitemShut
  {NoStop}%
\bibitem [{\citenamefont {Steinhardt}\ and\ \citenamefont
  {Turok}(2002)}]{Steinhardt:2001st}%
  \BibitemOpen
  \bibfield  {author} {\bibinfo {author} {\bibfnamefont {P.~J.}\ \bibnamefont
  {Steinhardt}}\ and\ \bibinfo {author} {\bibfnamefont {N.}~\bibnamefont
  {Turok}},\ }\href {\doibase 10.1103/PhysRevD.65.126003} {\bibfield  {journal}
  {\bibinfo  {journal} {Phys. Rev. D}\ }\textbf {\bibinfo {volume} {65}},\
  \bibinfo {pages} {126003} (\bibinfo {year} {2002})},\ \Eprint
  {http://arxiv.org/abs/hep-th/0111098} {arXiv:hep-th/0111098} \BibitemShut
  {NoStop}%
\bibitem [{\citenamefont {Brandenberger}\ and\ \citenamefont
  {Vafa}(1989)}]{Brandenberger:1988aj}%
  \BibitemOpen
  \bibfield  {author} {\bibinfo {author} {\bibfnamefont {R.~H.}\ \bibnamefont
  {Brandenberger}}\ and\ \bibinfo {author} {\bibfnamefont {C.}~\bibnamefont
  {Vafa}},\ }\href {\doibase 10.1016/0550-3213(89)90037-0} {\bibfield
  {journal} {\bibinfo  {journal} {Nucl. Phys. B}\ }\textbf {\bibinfo {volume}
  {316}},\ \bibinfo {pages} {391} (\bibinfo {year} {1989})}\BibitemShut
  {NoStop}%
\bibitem [{\citenamefont {Tseytlin}\ and\ \citenamefont
  {Vafa}(1992)}]{Tseytlin:1991xk}%
  \BibitemOpen
  \bibfield  {author} {\bibinfo {author} {\bibfnamefont {A.~A.}\ \bibnamefont
  {Tseytlin}}\ and\ \bibinfo {author} {\bibfnamefont {C.}~\bibnamefont
  {Vafa}},\ }\href {\doibase 10.1016/0550-3213(92)90327-8} {\bibfield
  {journal} {\bibinfo  {journal} {Nucl. Phys. B}\ }\textbf {\bibinfo {volume}
  {372}},\ \bibinfo {pages} {443} (\bibinfo {year} {1992})},\ \Eprint
  {http://arxiv.org/abs/hep-th/9109048} {arXiv:hep-th/9109048} \BibitemShut
  {NoStop}%
\bibitem [{\citenamefont {Kripfganz}\ and\ \citenamefont
  {Perlt}(1988)}]{Kripfganz:1987rh}%
  \BibitemOpen
  \bibfield  {author} {\bibinfo {author} {\bibfnamefont {J.}~\bibnamefont
  {Kripfganz}}\ and\ \bibinfo {author} {\bibfnamefont {H.}~\bibnamefont
  {Perlt}},\ }\href {\doibase 10.1088/0264-9381/5/3/006} {\bibfield  {journal}
  {\bibinfo  {journal} {Class. Quant. Grav.}\ }\textbf {\bibinfo {volume}
  {5}},\ \bibinfo {pages} {453} (\bibinfo {year} {1988})}\BibitemShut {NoStop}%
\bibitem [{\citenamefont {Boyle}\ \emph {et~al.}(2004)\citenamefont {Boyle},
  \citenamefont {Steinhardt},\ and\ \citenamefont {Turok}}]{Boyle:2003km}%
  \BibitemOpen
  \bibfield  {author} {\bibinfo {author} {\bibfnamefont {L.~A.}\ \bibnamefont
  {Boyle}}, \bibinfo {author} {\bibfnamefont {P.~J.}\ \bibnamefont
  {Steinhardt}}, \ and\ \bibinfo {author} {\bibfnamefont {N.}~\bibnamefont
  {Turok}},\ }\href {\doibase 10.1103/PhysRevD.69.127302} {\bibfield  {journal}
  {\bibinfo  {journal} {Phys. Rev. D}\ }\textbf {\bibinfo {volume} {69}},\
  \bibinfo {pages} {127302} (\bibinfo {year} {2004})},\ \Eprint
  {http://arxiv.org/abs/hep-th/0307170} {arXiv:hep-th/0307170} \BibitemShut
  {NoStop}%
\bibitem [{\citenamefont {Brandenberger}\ \emph {et~al.}(2007)\citenamefont
  {Brandenberger}, \citenamefont {Nayeri}, \citenamefont {Patil},\ and\
  \citenamefont {Vafa}}]{Brandenberger:2006xi}%
  \BibitemOpen
  \bibfield  {author} {\bibinfo {author} {\bibfnamefont {R.~H.}\ \bibnamefont
  {Brandenberger}}, \bibinfo {author} {\bibfnamefont {A.}~\bibnamefont
  {Nayeri}}, \bibinfo {author} {\bibfnamefont {S.~P.}\ \bibnamefont {Patil}}, \
  and\ \bibinfo {author} {\bibfnamefont {C.}~\bibnamefont {Vafa}},\ }\href
  {\doibase 10.1103/PhysRevLett.98.231302} {\bibfield  {journal} {\bibinfo
  {journal} {Phys. Rev. Lett.}\ }\textbf {\bibinfo {volume} {98}},\ \bibinfo
  {pages} {231302} (\bibinfo {year} {2007})},\ \Eprint
  {http://arxiv.org/abs/hep-th/0604126} {arXiv:hep-th/0604126} \BibitemShut
  {NoStop}%
\bibitem [{\citenamefont {Garcia}\ \emph {et~al.}(2017)\citenamefont {Garcia},
  \citenamefont {Mambrini}, \citenamefont {Olive},\ and\ \citenamefont
  {Peloso}}]{Garcia:2017tuj}%
  \BibitemOpen
  \bibfield  {author} {\bibinfo {author} {\bibfnamefont {M.~A.~G.}\
  \bibnamefont {Garcia}}, \bibinfo {author} {\bibfnamefont {Y.}~\bibnamefont
  {Mambrini}}, \bibinfo {author} {\bibfnamefont {K.~A.}\ \bibnamefont {Olive}},
  \ and\ \bibinfo {author} {\bibfnamefont {M.}~\bibnamefont {Peloso}},\ }\href
  {\doibase 10.1103/PhysRevD.96.103510} {\bibfield  {journal} {\bibinfo
  {journal} {Phys. Rev. D}\ }\textbf {\bibinfo {volume} {96}},\ \bibinfo
  {pages} {103510} (\bibinfo {year} {2017})},\ \Eprint
  {http://arxiv.org/abs/1709.01549} {arXiv:1709.01549 [hep-ph]} \BibitemShut
  {NoStop}%
\bibitem [{\citenamefont {Chen}\ and\ \citenamefont
  {Kang}(2018)}]{Chen:2017kvz}%
  \BibitemOpen
  \bibfield  {author} {\bibinfo {author} {\bibfnamefont {S.-L.}\ \bibnamefont
  {Chen}}\ and\ \bibinfo {author} {\bibfnamefont {Z.}~\bibnamefont {Kang}},\
  }\href {\doibase 10.1088/1475-7516/2018/05/036} {\bibfield  {journal}
  {\bibinfo  {journal} {JCAP}\ }\textbf {\bibinfo {volume} {05}},\ \bibinfo
  {pages} {036} (\bibinfo {year} {2018})},\ \Eprint
  {http://arxiv.org/abs/1711.02556} {arXiv:1711.02556 [hep-ph]} \BibitemShut
  {NoStop}%
\bibitem [{\citenamefont {Bernal}\ \emph {et~al.}(2019)\citenamefont {Bernal},
  \citenamefont {Elahi}, \citenamefont {Maldonado},\ and\ \citenamefont
  {Unwin}}]{Bernal:2019mhf}%
  \BibitemOpen
  \bibfield  {author} {\bibinfo {author} {\bibfnamefont {N.}~\bibnamefont
  {Bernal}}, \bibinfo {author} {\bibfnamefont {F.}~\bibnamefont {Elahi}},
  \bibinfo {author} {\bibfnamefont {C.}~\bibnamefont {Maldonado}}, \ and\
  \bibinfo {author} {\bibfnamefont {J.}~\bibnamefont {Unwin}},\ }\href
  {\doibase 10.1088/1475-7516/2019/11/026} {\bibfield  {journal} {\bibinfo
  {journal} {JCAP}\ }\textbf {\bibinfo {volume} {11}},\ \bibinfo {pages} {026}
  (\bibinfo {year} {2019})},\ \Eprint {http://arxiv.org/abs/1909.07992}
  {arXiv:1909.07992 [hep-ph]} \BibitemShut {NoStop}%
\bibitem [{\citenamefont {Garcia}\ and\ \citenamefont
  {Amin}(2018)}]{Garcia:2018wtq}%
  \BibitemOpen
  \bibfield  {author} {\bibinfo {author} {\bibfnamefont {M.~A.~G.}\
  \bibnamefont {Garcia}}\ and\ \bibinfo {author} {\bibfnamefont {M.~A.}\
  \bibnamefont {Amin}},\ }\href {\doibase 10.1103/PhysRevD.98.103504}
  {\bibfield  {journal} {\bibinfo  {journal} {Phys. Rev. D}\ }\textbf {\bibinfo
  {volume} {98}},\ \bibinfo {pages} {103504} (\bibinfo {year} {2018})},\
  \Eprint {http://arxiv.org/abs/1806.01865} {arXiv:1806.01865 [hep-ph]}
  \BibitemShut {NoStop}%
\bibitem [{\citenamefont {Freese}\ \emph {et~al.}(2024)\citenamefont {Freese},
  \citenamefont {Montefalcone},\ and\ \citenamefont {Shams
  Es~Haghi}}]{Freese:2024ogj}%
  \BibitemOpen
  \bibfield  {author} {\bibinfo {author} {\bibfnamefont {K.}~\bibnamefont
  {Freese}}, \bibinfo {author} {\bibfnamefont {G.}~\bibnamefont
  {Montefalcone}}, \ and\ \bibinfo {author} {\bibfnamefont {B.}~\bibnamefont
  {Shams Es~Haghi}},\ }\href {\doibase 10.1103/PhysRevLett.133.211001}
  {\bibfield  {journal} {\bibinfo  {journal} {Phys. Rev. Lett.}\ }\textbf
  {\bibinfo {volume} {133}},\ \bibinfo {pages} {211001} (\bibinfo {year}
  {2024})},\ \Eprint {http://arxiv.org/abs/2401.17371} {arXiv:2401.17371
  [hep-ph]} \BibitemShut {NoStop}%
\bibitem [{\citenamefont {Wang}\ \emph {et~al.}(2025)\citenamefont {Wang},
  \citenamefont {Jia}, \citenamefont {Chen},\ and\ \citenamefont
  {Tang}}]{Wang:2025duy}%
  \BibitemOpen
  \bibfield  {author} {\bibinfo {author} {\bibfnamefont {Q.-Y.}\ \bibnamefont
  {Wang}}, \bibinfo {author} {\bibfnamefont {T.}~\bibnamefont {Jia}}, \bibinfo
  {author} {\bibfnamefont {P.-R.}\ \bibnamefont {Chen}}, \ and\ \bibinfo
  {author} {\bibfnamefont {Y.}~\bibnamefont {Tang}},\ }\href {\doibase
  10.1103/jdkz-8z54} {\bibfield  {journal} {\bibinfo  {journal} {Phys. Rev. D}\
  }\textbf {\bibinfo {volume} {112}},\ \bibinfo {pages} {043508} (\bibinfo
  {year} {2025})},\ \Eprint {http://arxiv.org/abs/2504.13147} {arXiv:2504.13147
  [astro-ph.CO]} \BibitemShut {NoStop}%
\bibitem [{\citenamefont {Ghiglieri}\ and\ \citenamefont
  {Laine}(2015)}]{Ghiglieri:2015nfa}%
  \BibitemOpen
  \bibfield  {author} {\bibinfo {author} {\bibfnamefont {J.}~\bibnamefont
  {Ghiglieri}}\ and\ \bibinfo {author} {\bibfnamefont {M.}~\bibnamefont
  {Laine}},\ }\href {\doibase 10.1088/1475-7516/2015/07/022} {\bibfield
  {journal} {\bibinfo  {journal} {JCAP}\ }\textbf {\bibinfo {volume} {07}},\
  \bibinfo {pages} {022} (\bibinfo {year} {2015})},\ \Eprint
  {http://arxiv.org/abs/1504.02569} {arXiv:1504.02569 [hep-ph]} \BibitemShut
  {NoStop}%
\bibitem [{\citenamefont {Ghiglieri}\ \emph {et~al.}(2020)\citenamefont
  {Ghiglieri}, \citenamefont {Jackson}, \citenamefont {Laine},\ and\
  \citenamefont {Zhu}}]{Ghiglieri:2020mhm}%
  \BibitemOpen
  \bibfield  {author} {\bibinfo {author} {\bibfnamefont {J.}~\bibnamefont
  {Ghiglieri}}, \bibinfo {author} {\bibfnamefont {G.}~\bibnamefont {Jackson}},
  \bibinfo {author} {\bibfnamefont {M.}~\bibnamefont {Laine}}, \ and\ \bibinfo
  {author} {\bibfnamefont {Y.}~\bibnamefont {Zhu}},\ }\href {\doibase
  10.1007/JHEP07(2020)092} {\bibfield  {journal} {\bibinfo  {journal} {JHEP}\
  }\textbf {\bibinfo {volume} {07}},\ \bibinfo {pages} {092} (\bibinfo {year}
  {2020})},\ \Eprint {http://arxiv.org/abs/2004.11392} {arXiv:2004.11392
  [hep-ph]} \BibitemShut {NoStop}%
\bibitem [{\citenamefont {Ringwald}\ \emph {et~al.}(2021)\citenamefont
  {Ringwald}, \citenamefont {Sch{\"u}tte-Engel},\ and\ \citenamefont
  {Tamarit}}]{Ringwald:2020ist}%
  \BibitemOpen
  \bibfield  {author} {\bibinfo {author} {\bibfnamefont {A.}~\bibnamefont
  {Ringwald}}, \bibinfo {author} {\bibfnamefont {J.}~\bibnamefont
  {Sch{\"u}tte-Engel}}, \ and\ \bibinfo {author} {\bibfnamefont
  {C.}~\bibnamefont {Tamarit}},\ }\href {\doibase
  10.1088/1475-7516/2021/03/054} {\bibfield  {journal} {\bibinfo  {journal}
  {JCAP}\ }\textbf {\bibinfo {volume} {03}},\ \bibinfo {pages} {054} (\bibinfo
  {year} {2021})},\ \Eprint {http://arxiv.org/abs/2011.04731} {arXiv:2011.04731
  [hep-ph]} \BibitemShut {NoStop}%
\bibitem [{\citenamefont {Aggarwal}\ \emph {et~al.}(2021)\citenamefont
  {Aggarwal} \emph {et~al.}}]{Aggarwal:2020olq}%
  \BibitemOpen
  \bibfield  {author} {\bibinfo {author} {\bibfnamefont {N.}~\bibnamefont
  {Aggarwal}} \emph {et~al.},\ }\href {\doibase 10.1007/s41114-021-00032-5}
  {\bibfield  {journal} {\bibinfo  {journal} {Living Rev. Rel.}\ }\textbf
  {\bibinfo {volume} {24}},\ \bibinfo {pages} {4} (\bibinfo {year} {2021})},\
  \Eprint {http://arxiv.org/abs/2011.12414} {arXiv:2011.12414 [gr-qc]}
  \BibitemShut {NoStop}%
\bibitem [{\citenamefont {Aggarwal}\ \emph
  {et~al.}(2025{\natexlab{b}})\citenamefont {Aggarwal} \emph
  {et~al.}}]{Aggarwal:2025noe}%
  \BibitemOpen
  \bibfield  {author} {\bibinfo {author} {\bibfnamefont {N.}~\bibnamefont
  {Aggarwal}} \emph {et~al.},\ }\href@noop {} {\  (\bibinfo {year}
  {2025}{\natexlab{b}})},\ \Eprint {http://arxiv.org/abs/2501.11723}
  {arXiv:2501.11723 [gr-qc]} \BibitemShut {NoStop}%
\bibitem [{\citenamefont {Herman}\ \emph {et~al.}(2023)\citenamefont {Herman},
  \citenamefont {Lehoucq},\ and\ \citenamefont {F{\'{u}}zfa}}]{Herman:2022fau}%
  \BibitemOpen
  \bibfield  {author} {\bibinfo {author} {\bibfnamefont {N.}~\bibnamefont
  {Herman}}, \bibinfo {author} {\bibfnamefont {L.}~\bibnamefont {Lehoucq}}, \
  and\ \bibinfo {author} {\bibfnamefont {A.}~\bibnamefont {F{\'{u}}zfa}},\
  }\href {\doibase 10.1103/PhysRevD.108.124009} {\bibfield  {journal} {\bibinfo
   {journal} {Phys. Rev. D}\ }\textbf {\bibinfo {volume} {108}},\ \bibinfo
  {pages} {124009} (\bibinfo {year} {2023})},\ \Eprint
  {http://arxiv.org/abs/2203.15668} {arXiv:2203.15668 [gr-qc]} \BibitemShut
  {NoStop}%
\bibitem [{\citenamefont {Bernal}\ and\ \citenamefont
  {Xu}(2025)}]{Bernal:2024jim}%
  \BibitemOpen
  \bibfield  {author} {\bibinfo {author} {\bibfnamefont {N.}~\bibnamefont
  {Bernal}}\ and\ \bibinfo {author} {\bibfnamefont {Y.}~\bibnamefont {Xu}},\
  }\href {\doibase 10.1007/JHEP01(2025)137} {\bibfield  {journal} {\bibinfo
  {journal} {JHEP}\ }\textbf {\bibinfo {volume} {01}},\ \bibinfo {pages} {137}
  (\bibinfo {year} {2025})},\ \Eprint {http://arxiv.org/abs/2410.21385}
  {arXiv:2410.21385 [hep-ph]} \BibitemShut {NoStop}%
\bibitem [{\citenamefont {Montefalcone}\ \emph {et~al.}(2025)\citenamefont
  {Montefalcone}, \citenamefont {Shams Es~Haghi}, \citenamefont {Xu},\ and\
  \citenamefont {Freese}}]{Montefalcone:2025gxx}%
  \BibitemOpen
  \bibfield  {author} {\bibinfo {author} {\bibfnamefont {G.}~\bibnamefont
  {Montefalcone}}, \bibinfo {author} {\bibfnamefont {B.}~\bibnamefont {Shams
  Es~Haghi}}, \bibinfo {author} {\bibfnamefont {T.}~\bibnamefont {Xu}}, \ and\
  \bibinfo {author} {\bibfnamefont {K.}~\bibnamefont {Freese}},\ }\href
  {\doibase 10.1103/rnvb-t4lx} {\bibfield  {journal} {\bibinfo  {journal}
  {Phys. Rev. D}\ }\textbf {\bibinfo {volume} {112}},\ \bibinfo {pages}
  {063556} (\bibinfo {year} {2025})},\ \Eprint
  {http://arxiv.org/abs/2507.08739} {arXiv:2507.08739 [hep-ph]} \BibitemShut
  {NoStop}%
\end{thebibliography}%
\end{document}